\documentclass[aps,pra]{revtex4}
\begin{document}

\title{Correct mutual information, quantum bit error rate and secure
transmission efficiency in Wojcik's eavesdropping scheme on
ping-pong protocol
\thanks{E-mail: Zhangzj@wipm.ac.cn}}

\author{
Zhanjun Zhang  \\
{\footnotesize  Wuhan Institute of Physics and Mathematics, The
Chinese Academy of Sciences, Wuhan 430071, China \\
E-mail: Zhangzj@wipm.ac.cn} }

\date{\today}

\maketitle

Recently W\'{o}jcik has analyzed the security of the 'ping-pong'
quantum communication protocol[1] in the case of considerable
quantum channel losses and accordingly an undetectable
eavesdropping scheme and possible improvements on the 'ping-pong'
protocol are proposed [2]. This is true. However, in his paper,
there exist some mistakes, i.e., the mutual information $I$,
quantum bit error rate (QBER) and secure transmission efficiency
are incorrect. In this paper, I would like to point out them and
correct them.

First, I would like to present a simple question and an obvious
fact:

(a) In Wojcik's paper, Bob's QBER (quantum bits error rate) is
precisely 1/4. Then one would like to ask how many wrong bits in
Bob's bits provided that the total bit number N is odd or 2n+2.
Assuming N=201,202,203, whether the corresponding answers are
50.25,50.50,50.75 respectively? If so, how ridiculous they are for
the numbers of the wrong bits are noninteger. It is the wrong
QBER=1/4 that leads to the ridiculous results.

(b) According to Wojcik's figure 4, one can find when the channel
transmission efficiency is zero, both Eve and Bob still can get
informations from Alice and Eve's is larger than Bob's. This is
like that, when I call you and speak aloud for a quite long time
disregarding the status of the phone, although the phoneline is
cutted at the very beginning, you still can get some information
from me. Another ridiculous result!

All these have obviously shown that there are mistakes in Wojcik's
paper.

I strongly suspect Wojick's conclusion (i.e., Eve's mutual
information with Alice's is not decreased for she exactly knows
whether the symmetry operation is performed or not, while Bob's
mutual information with Alice's is decreased for he does not know
whether the symmetry operation is performed. Alternatively, in
Wojcik's original words, 'It should be emphasized that
symmetrization disturbs the communication between Alice and Bob in
such a way that mutual information is reduced while the QBER is
not affected'). Here I offer a verification of my suspicion via
simple examples. Additionally, in these examples the doubtful
QBER=1/4 question can also be addressed.

Let 'u' denote Eve's attack without the symmetry operation and 's'
be Eve's attack with the symmetry operation. Suppose Alice's bit
be '1001'. If Bob gets '1000' after his measurements, one can
easily work out the mutual information between Bob and Alice
$I_{AB}=0.311$ and Bob's QBER=1/4. The calculations are completely
determined by Bob's bits and Alice's bits disregarding the
intermediate process by Eve. In fact, after any one of the
'uuuu','susu','ssuu' or  'ussu' attacks by Eve, it is possible for
Bob to get '1000'. So one can easily find that after the symmetry
operations (e.g.,'susu', 'ssuu' and 'ussu') the mutual information
between Bos and Alice is not reduced at all. This indicates that
Wojcik's correspoding conclusion is wrong. Although the QBER in
the above example is 1/4, in fact, Bob's QBER is not always 1/4 at
all. This can be seen from the following example. Also suppose
Alice's bits be '1001'. It is possible for Bob to get '1001' after
any one of the 'uuuu', 'ssss', 'suus', 'susu', 'ssuu', 'usus',
ussu' or 'uuss' attacks by Eve. In any one of these cases, the
$I_{AB}=1$ while the QBER=0. This example also verifies that
Wojcik's conclusion is wrong. In fact, it is possible for Bob to
get different bits after his measurements. Generally speaking,
different bits correspond to different mutual information and
different QBER. This conclusion is different from Wojcik's given
mutual information and given QBER.

I would like to say the essential reason which leads to Wojcik's
is the inappropriate use of the mutual information. He has made a
confusion of the mutual information of single-bit-gain and the
mutual information of multi-bits-gain. Also he should notice in
reality the number of Alice's bits is finite.

As for discussions on the secure transmission efficiency, one can
also see my manuscript quant-ph-0402022.

In Wojcik's paper, according to the analysis on the mutual
information $I$ as a function of the transmission efficiency
$\eta$ of the quantum channel, W\'{o}jcik concludes that the
'ping-pong' protocol is not secure for $\eta < 60\%$. Here I would
like to point out that this conclusion is not reliable.

The mutual information in Wojcik's paper is only and essentially a
single-time- statistic mutual information. This physical quantity
is used inappropriately in Wojcik's paper to stand for the mutual
information of the whole transmissions (multi times),
alternatively, the mutual informations of the whole transmissions
(multi times) are not worked out appropriately ( It is not
suitable to calculate the mutual informations by using the
so-called joint probability distributions as the equation 7 in
W\'{o}jcik's paper). Since the number $N$ of Alice's transmitted
bits is finite in reality, generally speaking, after the whole
transmissions, it is quite possible that APD (probability
distributions which can be and should be extracted from all the
Alice's, Bob's and Eve's bits) are different from SPD (probability
distributions suitable for the single bit gain). The larger $N$,
the more possibly APD are close to SPD. Only when $N$ is infinite,
APD are equal to SPD.

Consider the case that $\eta \leq 50\%$ (i.e., Eve can attack all
the bits). Suppose Alice's transmitted and attacked bits be
$'1011000101\dots'$. When Alice sends the first bit '1', according
to W\'{o}jcik's scheme, it is {\it possible} for Bob to get '1'
and for Eve to get '0' by measurements. Also it is {\it possible}
for Bob to get '0' and for Eve to get '1'. Certainly there are
other possibilities. Similar results occur for other transmitted
bits. So after Alice transmits the bits one by one, it is {\it
possible} for Bob to get $'1011000101\dots'$. This means that it
is {\it possible} for Bob to get same bits with Alice's provided
that $N$ is finite (In reality $N$ should be finite). This denies
W\'{o}jcik's conclusion that $I_{AE}$ is {\it always} bigger than
$I_{AB}$ when $\eta \leq 50\%$ (cf. figure 4 in W\'{o}jcik's
paper). Additionally, it is unimaginable that according to the
figure 4 both $I_{AE}$ and $I_{AB}$ are nonzero and $I_{AE}>
I_{AB}$ when $\eta=0$. Of course, similarly, it is also {\it
possible} for Eve to get the same bits with Alice's when $N$ is
finite. In the case that $N$ is infinite $I_{AE}=I_{AB}$ can be
arrived at. All these mean that the ping-pong protocol is really
not secure when $\eta \leq 50\%$. Still in the case that $N$ is
infinite, when $\eta$ is larger than $50\%$, then only a fraction
of all the transmitted bits can be attacked by Eve and accordingly
$I_{AB}>I_{AE}$. The forking point is $\eta=50\%$. This means that
the corresponding conclusion in W\'{o}jcik's paper is wrong. In
reality $N$ should be finite. In this case, for all the attacked
bits, it is {\it possible} for Eve to get the same ones with
Alice's. Therefore, if a condition that the number of the Bob's
correct bits is not less than the number of Eve's correct bits is
required, it is easy to work out that $\eta$ should be not less
than $75\%$. This means that in reality the ping-pong protocol is
not secure for $\eta \leq 75\%$.

This work is supported by the National Natural Science Foundation
of China under Grant Nos. 10304022.

\noindent [1] K. Bostr\"{o}m and T. Felbinger, Phys. Rev. Lett.
{\bf 89}, 187902(2002).

\noindent [2] Antoni W\'{o}jcik, Phys. Rev. Lett., {\bf 90},
157901(2003).

\end{document}